\documentclass[aps,pra,twocolumn,superscriptaddress]{revtex4-1}

\usepackage{graphicx}
\usepackage{hyperref}
\usepackage{amsmath}
\usepackage{amssymb}
\usepackage{epstopdf}
\usepackage{multirow}
\usepackage{threeparttable}
\usepackage{xcolor}

\begin{document}

\title{High-speed mid-infrared imaging via nonlinear multiplexed detection}

\author{Ruiyang Qin}
\affiliation{State Key Laboratory of Precision Spectroscopy, and Hainan Institute, East China Normal University, Shanghai 200062, China}

\author{Kun Huang}
\email{khuang@lps.ecnu.edu.cn}
\affiliation{State Key Laboratory of Precision Spectroscopy, and Hainan Institute, East China Normal University, Shanghai 200062, China}
\affiliation{Chongqing Key Laboratory of Precision Optics, Chongqing Institute of East China Normal University, Chongqing 401121, China}
\affiliation{Collaborative Innovation Center of Extreme Optics, Shanxi University, Taiyuan, Shanxi 030006, China}

\author{Min Peng}
\affiliation{State Key Laboratory of Precision Spectroscopy, and Hainan Institute, East China Normal University, Shanghai 200062, China}

\author{Jianan Fang}
\affiliation{State Key Laboratory of Precision Spectroscopy, and Hainan Institute, East China Normal University, Shanghai 200062, China}

\author{Ben Sun}
\affiliation{State Key Laboratory of Precision Spectroscopy, and Hainan Institute, East China Normal University, Shanghai 200062, China}

\author{Zhengru Guo}
\affiliation{University of Shanghai for Science and Technology, School of Optical Electrical and Computer Engineering, Shanghai, China}

\author{Heping Zeng}
\email{hpzeng@phy.ecnu.edu.cn}
\affiliation{State Key Laboratory of Precision Spectroscopy, and Hainan Institute, East China Normal University, Shanghai 200062, China}
\affiliation{Chongqing Key Laboratory of Precision Optics, Chongqing Institute of East China Normal University, Chongqing 401121, China}
\affiliation{Shanghai Research Center for Quantum Sciences, Shanghai 201315, China}
\affiliation{Chongqing Institute for Brain and Intelligence, Guangyang Bay Laboratory, Chongqing, 400064, China}

\begin{abstract}
High-speed mid-infrared (MIR) videography constitutes an enabling tool to monitor and analyze various dynamics in scientific research and industrial applications, such as combustion diagnostics, explosion reactions, photosynthetic tracking, and thermal surveillance. However, the frame rate of conventional MIR imagers is typically limited by readout electronics and detection sensitivity, especially for large spatial formats with massive pixels. Here, we devise and implement a high-speed MIR upconversion imaging system based on time-multiplexed nonlinear structured pumping. Specifically, the dynamic infrared scene is optically gated by a sequence of spatially periodical pump patterns in a nonlinear crystal, which facilitates both rapid temporal encryption and sensitive upconversion detection. Then, the upconverted frames are superimposed onto a silicon camera within a single exposure, thus resulting in a multiplexed snapshot in the spatial-frequency domain. Finally, the sub-exposure images, corresponding to distinct transient events, can be computationally deciphered and reconstructed by the frequency recognition algorithm based on band-pass filtering and Fourier transform operations. The achieved frame rate is tenfold boosted to 10,000 frames per second without compromising the megapixel spatial format, which allows continuous real-time MIR videography at high speed and high definition. The presented approach could be readily extended to far-infrared or terahertz spectral regions, with an aim of performing high-throughput and high-sensitivity observation of transient phenomena with high temporal complexity.
\end{abstract}

\maketitle

\section{Introduction}
High-speed optical imaging is a powerful technique to capture transient events and fast phenomena, which has proven pivotal in a wide range of applications from shockwave monitoring, plasma observation, combustion analysis to aerospace engineering and collision inspection \cite{Liang2018Optica, Zeng2023US, Qi2020AP}. Imaging systems with high spatiotemporal resolution across a large field of view are highly desirable to meet these demands \cite{Yao2024LPR, Liang2024Book}. To date, cutting-edge fast cameras based on in-situ storage image sensors have been developed to demonstrate high frame rates over million frames per second \cite{Etoh20213Sensor}, albeit with technical limitations in pixel counts, sequence depth, and light throughput. Notably, the data acquisition rate is ultimately limited by the processing time during the charge transfer and storage, which imposes an upper limit on the attainable speed of a high-speed camera \cite{Qi2020AP, Yao2024LPR}. To further boost the imaging speed, the time-division framing photographic technique has been proposed to perform a multiplexing operation within a single-shot exposure \cite{Liang2024Book}, where the temporal slice for each transient event is encoded into intrinsic dimensions of the optical field, such as positions \cite{Wang2014AO}, angles \cite{Li2014NC}, and wavelengths \cite{Nakagawa2014NP}. Moreover, in combination with advanced algorithms of compressive sensing \cite{Gao2014Nature, Wang2024LAM} or machine learning \cite{Liu2022NC, Chen2024Photonix}, the computational ultrafast imaging framework has emerged to significantly augment the frame rate, which allows the recording of non-repeatable or destructive events.

However, the aforementioned high-speed imaging techniques have been typically demonstrated in the visible or near-infrared (NIR) region due to the availability of sensitive detectors and fast modulators \cite{Liang2024Book}. Nowadays, there are increasing interests in extending the operation wavelength into the mid-infrared (MIR) regime, pertinent to appealing applications that need to acquire thermal and chemical information at high frame rates \cite{Xie2017Optica}. Representative examples for the high-speed MIR videography include heat transfer analysis, combustion diagnostics, chemical reactions, and thermal surveillance \cite{Vodopyanov2020Book}. In comparison to the silicon-based cameras, conventional MIR imagers based on narrow-bandgap semiconductors or pyroelectric bolometers exhibit much lower frame rates due to the limitations on the on-chip storage and readout speed \cite{Razeghi2014RPP}. For instance, state-of-the-art MIR cameras based on indium antimonide (InSb) are specified with a maximum full frame rate of 180 Hz for a megapixel spatial format \cite{Telops}. Moreover, an improved readout rate is usually permitted at the price of a smaller focal plane array, due to the trade-off between pixel density and the overall response time of the arrayed detectors. In addition, MIR imagers have long suffered from severe dark currents and thermal noises at room temperature \cite{Wang2019Small}, which often require active cooling to operate at the cryogenic condition. Therefore, it becomes crucial to enhance the detection sensitivity to approach high-speed MIR imaging with a high signal-to-noise ratio (SNR), as the number of photons collected per frame decreases with shorter integration times.

In this context, tremendous efforts have been dedicated to developing the so-called frequency upconversion imaging technology, where the information encoded in the MIR radiation is transferred to the replica at short wavelengths for leveraging the high-performance sensors with high efficiency, low noise, and fast response \cite{Barh2019AOP, Dam2012NP, Huang2022NC}. Such an intermediate step can be realized by resorting to nonlinear media based on bulky crystals \cite{Wang2023NC, Ge2023PRAppl, Mrejen2020LPR}, thin films \cite{Zhu2022Optica} and resonant metasurfaces \cite{Molina2024AM, Zheng2023OEA}. Alternatively, the detector material itself can be used to offer the optical nonlinearity based on non-degenerate two-photon absorption (TPA) in large-bandgap semiconductors \cite{Fishman2011NP, Potma2021APLP, Fang2021IEEE}. The upconversion strategy favors sensitive and fast MIR sensing at room temperature, hence leading to demonstrations of high-speed grating spectroscopy \cite{Zheng2023LPR, Rodrigo2021LPR}, real-time volumetric tomography \cite{Rehain2020NC, Fang2023LSA, Israelsen2019LSA}, and video-rate hyperspectral imaging \cite{Fang2024NC, Junaid2019Optica, Knez2022SA, Zhao2023NC}. However, in these instantiations, the available refreshing rate for the image data acquisition is still determined by the frame rate of used cameras. Notably, a binning operation permits a higher frame rate, albeit with a reduced number of active pixels \cite{Huang2022NC, Potma2021APLP}. Therefore, it remains appealing to develop novel techniques to implement high-speed and high-definition MIR imaging, with an aim to overcome the speed limitations of current large-format cameras.

Here, we propose and implement a high-speed MIR upconversion imaging system based on time-multiplexed nonlinear structured pumping. In our methodology, the incident dynamic scene is optically gated by a sequence of spatially modulated pump patterns in a nonlinear crystal, which facilitates both rapid temporal encoding and sensitive upconversion detection. The multiplexed information in the upconverted field is then recorded by a silicon camera within a single integration window. Each sub-exposure image is located in a distinct region at the Fourier domain. Finally, the transient frames are computationally extracted by performing the band-pass filtering and Fourier transform operations. The essence of the approach is to stack multiple frames into a single image to realize the snapshot videography, which enables one to break the physical limit on the frame rate of the camera. In our experiment, the MIR upconversion imaging speed is boosted by an order of magnitude to reach a frame rate of 10 kHz, while maintaining the megapixel spatial resolution. In this proof-of-principle demonstration, the achieved frame rate is dictated by the switching time of the digital micromirror device (DMD). Further improvements in the imaging speed are possible by resorting to faster pattern generation \cite{Kilcullen2022NC} or path-division configuration \cite{Ehn2017LSA}, which would pave a novel path toward the ultrafast MIR videography with a wide field of view.
 
\section{Basic principle}
The core of the high-speed MIR multiplexed imaging lies in the all-optical control of the MIR beam through the spatial modulation of the pump field. Such an indirect approach for the MIR modulation provides a solution to overcome the limited operation window for current mature spatial light modulators (SLMs) based on liquid crystals or micromirror arrays in the visible or NIR spectral region \cite{Wang2023NC}. Although the silver-coated mirrors in principle permit the operation at longer wavelengths, the beam steering efficacy is intrinsically restricted by parasitic diffraction phenomena, which in turn degrades the spatial modulation accuracy \cite{Edgar2019NP}. Here, the spatially dynamic pumping allows simultaneous functionalities of temporal aperture coding and nonlinear frequency upconversion. The involved nonlinear structured detection scheme differs from previous multiplexed image capturing approaches \cite{Wang2024LAM, Ehn2017LSA, Dong2023NC}, where the illumination beam is only subject to the spatial modulation without the wavelength conversion. Hence, the proposed architecture offers desirable features of high-fidelity modulation and high-sensitivity detection for the MIR radiation, which are two prerequisites for high-speed multiplexed imaging.

Specifically, the required nonlinear conversion is performed based on the second-order sum-frequency generation (SFG) process, where the MIR signal photon at an angular frequency of $\omega_s$ is spectrally converted to the upconversion replica with a higher frequency at $\omega_\text{up}$ under the pump field at $\omega_\text{p}$. The three angular frequencies satisfy the energy conservation as $\omega_\text{up}=\omega_\text{s}+\omega_\text{p}$. In the non-unsaturated regime, the SFG intensity is directly proportional to the product of the signal and pump intensities, as given by $I_\text{up}(x, y) \propto I_s(x, y) \times I_p(x, y)$ \cite{Wang2023NC}. This relation indicates that the spatial modulation of the pump is mathematically equivalent to modulate the signal. Consequently, the pump intensity pattern serves as a spatial transmission mask onto the MIR object image, and the filtered pattern is exactly mapped to the intensity profile of the upconverted field at a disparate wavelength.

\begin{figure*}[t!]
\includegraphics[width=0.82 \textwidth]{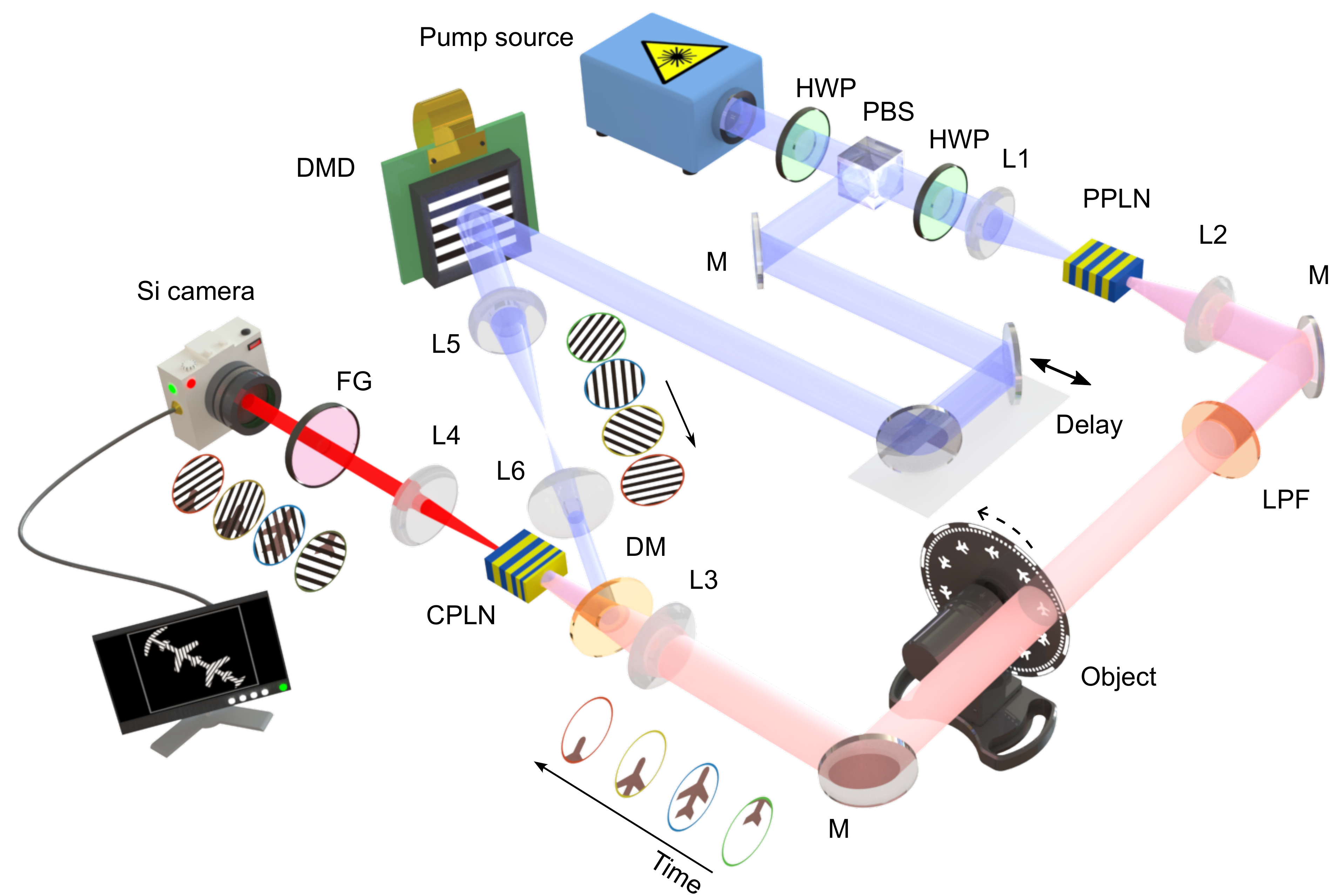}
\caption{Experimental setup of the time-multiplexed MIR upconversion imaging system. The involved laser source originates from a mode-locked ultrafast laser at 1030 nm. The output light is divided into two portions via a polarization beam splitter (PBS). The transmitted portion is injected into a periodically poled lithium niobate (PPLN) crystal to perform the optical parametric generation for preparing the MIR illumination source. The signal beam passes through a fast rotating disc carved with clear-path targets as the dynamic scene. The resulting object image is projected by a lens (L3) into the center of a chirped-poling lithium niobate (CPLN) crystal. Meanwhile, the laser pulses from the reflected portion is combined by a dichroic mirror (DM) into the nonlinear crystal, which serves as the synchronous pump to facilitate the sum-frequency generation. Notably, the pump beam is rapidly modulated by a programmed digital micromirror device (DMD), which delivers a sequence of spatially periodical patterns to implement the optical modulation on the MIR radiation. Consequently, the nonlinear structured mask favors both the temporal spatial encoding and frequency upconversion detection. After a group of spectral filters, the multiplexed upconversion image is captured by a high-definition silicon camera in a single cumulative exposure window. The sub-exposure images can be computationally reconstructed to reveal the transient events at a frame rate beyond the intrinsic capability of the camera. HWP: half-wave plate; M: silver mirror; LPF: long-pass filer; FG: filter group.}
\label{fig1}
\end{figure*}

To perform the multiplexing operation at the Fourier domain, a series of periodical patterns with distinct spatial frequencies are prepared as the pump source to shape the incident MIR radiation. For this work, a square wave is used to model the DMD modulation pattern. Without loss of generality, we consider a specific intensity distribution along the vertical direction with a spatial period of $T$, as expressed by
\begin{equation}
I_p(x) = I_p(x+T) = 
\begin{cases}
1\,, & |x| < T/4 \\
0\,, & T/4 \leq |x| \leq T/2 
\end{cases} \ .
\label{eq2}   
\end{equation}
General pump patterns $I_p^\text{(n)}(x, y)$ at arbitrary orientation angles and spatial periods can be derived by applying the rotation transformation at $\theta_n$ and designating proper values of $T_n$. By using the convolution theorem, the recorded upconversion image can be represented in the spatial frequency domain: $\tilde{I}_\text{up}(k_x, k_y) \propto \tilde{I}_\text{s}(k_x, k_y) \ast \tilde{I}_\text{p}(k_x, k_y)$, where $k_{x,y}$ denote spatial frequencies in the Cartesian coordinates, and the asterisk symbol represents the convolution operation. For a time-varying scene, the camera exposure window is segmented with the illumination duration of each modulation pattern, which ensures that each sub-exposure frame corresponds to a specific pattern. After integrating $N$ patterns, the multiplexed image in the Fourier domain is given by
\begin{equation}
\begin{aligned}
\tilde{I}_\text{up}(k_x, k_y) \propto & \sum_{n=1}^{N} \Big[ \tilde{I_s}(k_x, k_y) \ \ast  \\
& \sum_{m=-\infty}^{\infty} \frac{2 \sin (m k_0^\text{(n)} T_n/4)}{m} \delta[k^\text{(n)} - m k_0^\text{(n)}] \Big]
\label{eq3}
\end{aligned} \ ,
\end{equation}
where $k_0^\text{(n)} = 2\pi /T_n$ is the fundamental spatial frequency and $k^\text{(n)} = k_x \sin\theta _n + k_y \cos\theta _n $ for the $n$-th modulation pattern. It can be seen that the Dirac function $\delta$ produces replicas of the original image in the Fourier domain. Each copy of the image information is shifted to the location of higher-order harmonic components. Note that the first-order component is usually the focus of the image analysis due to the dominant magnitude in comparison to the higher-order ones. As a result, the multiplexed structured detection scheme can cast the temporal lapse into the spatial frequency shifts, which makes it possible for a single snapshot image to contain distinct time evolution \cite{Ehn2017LSA}. Each transient frame can be demultiplexed and recovered by performing numerical operations of band-pass frequency filtering and inverse Fourier transformation. The demodulation operation is analogous to the hyperdyne mixing with two orthogonal reference quadratures in the temporal domain \cite{Gragston2018AO}. In the presented method, the imaging speed is determined by the separation time between two sequential modulation patterns, which can overcome the intrinsic frame rate of the used camera. More details on the theoretical model of the nonlinear multiplexed imaging and the reconstruction procedure of the multiplexed images are presented in Supplementary Notes 2 and 3, respectively.

\section{Experimental setup}
Figure \ref{fig1} illustrates the experimental setup for the high-speed MIR imaging system based on the nonlinear multiplexed detection. The involved light source is from a mode-locked ultrafast laser at 1030 nm, which delivers a train of pulses with a duration of 5 ps at a repetition rate of 100 kHz. The laser output is split into two parts. One is injected into a periodically poled lithium niobate (PPLN) crystal for preparing the MIR signal source at 3.1 $\mu$m via the optical parametric generation. Then, the signal beam illuminates a dynamically changing scene that is emulated by a fast-moving object carved on a spinning wheel. The transmitted object image is relayed via a CaF$_2$ lens (LBTEK, MCX70610) into the center of a chirped-poling lithium niobate (CPLN) crystal. Meanwhile, the other part of the laser source after a delay line is used as the pump to perform the SFG. The CPLN is fabricated with linearly-ramping poling periods from 16 to 24 $\mu$m, which ensures quasi-phase matching for different incident angles \cite{Huang2022NC}. The extended acceptance angle favors high-resolution upconversion imaging in the 2f configuration \cite{Barh2019AOP}. Thanks to the coincidence-pumping architecture, the ultrafast pulsed pump provides high peak power to enhance the conversion efficiency, meanwhile suppressing the dark noise within a narrow temporal window. The SFG conversion efficiency is measured to be about 0.3\% at a pump power of 200 mW, which allows a sufficient SNR to capture MIR transient scenes. The efficiency can be further improved by increasing the pump intensity or employing a longer nonlinear crystal.

\begin{figure}[b!]
	\includegraphics[width=0.95 \columnwidth]{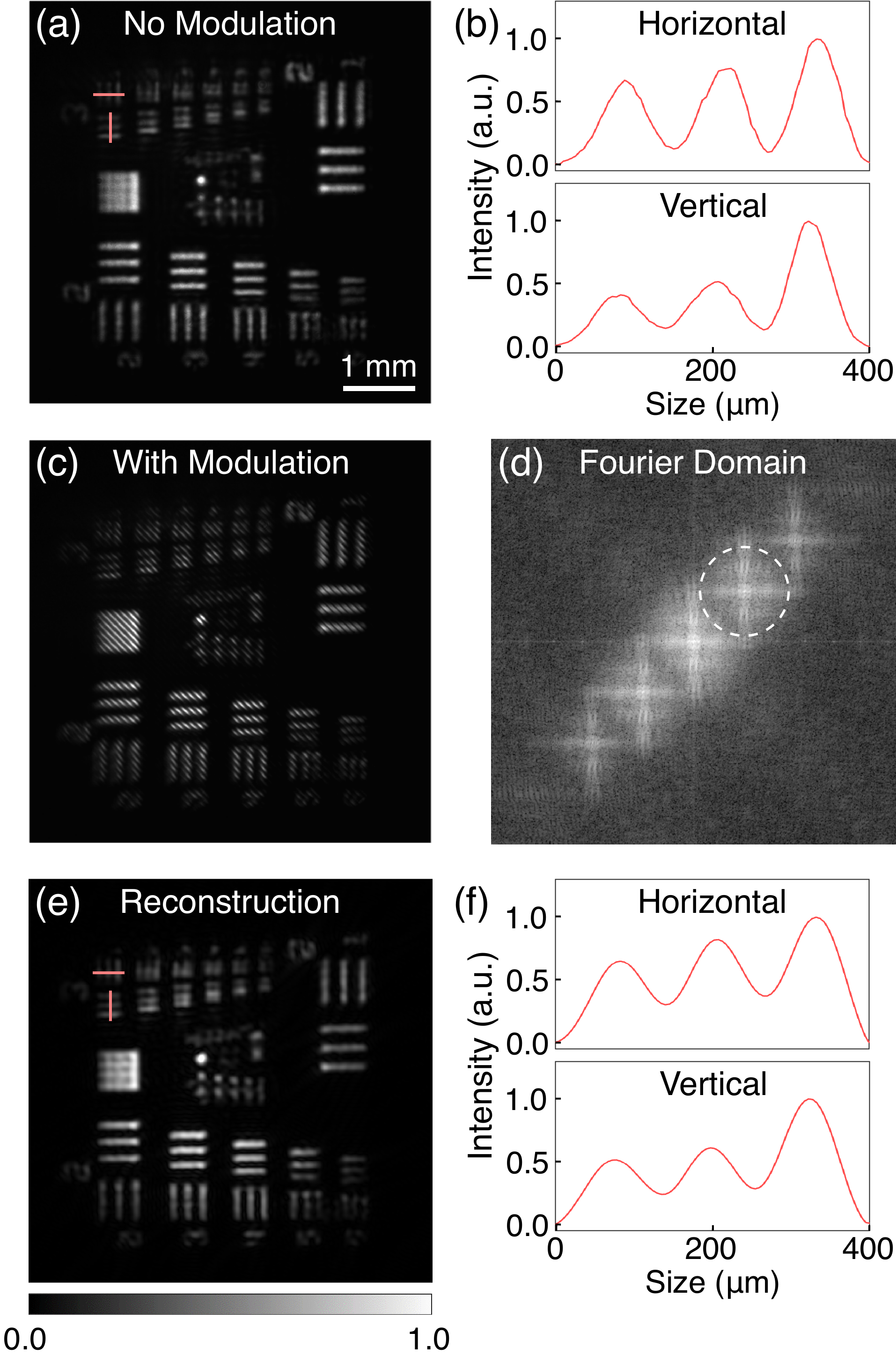}
	\caption{Characterization of the spatial resolution for the MIR imaging system. (a) Recorded upconversion image for a 1951 USAF resolution test chart without the spatial modulation. (b) Cross sections along the horizontal and vertical directions. (c) Representative upconversion image with a spatial modulation along 135$^\circ$. (d) Corresponding Fourier frequency distribution. Note that the dashed circle indicates the selected region for subsequent image reconstruction. A logarithmic operation is applied on the spectral magnitude to visualize the weak high-frequency components. (e) Demultiplexed image. (f) Cross sections along two orthogonal directions.}
	\label{fig2}
\end{figure}

\begin{figure*}[t!]
\includegraphics[width=0.58 \textwidth]{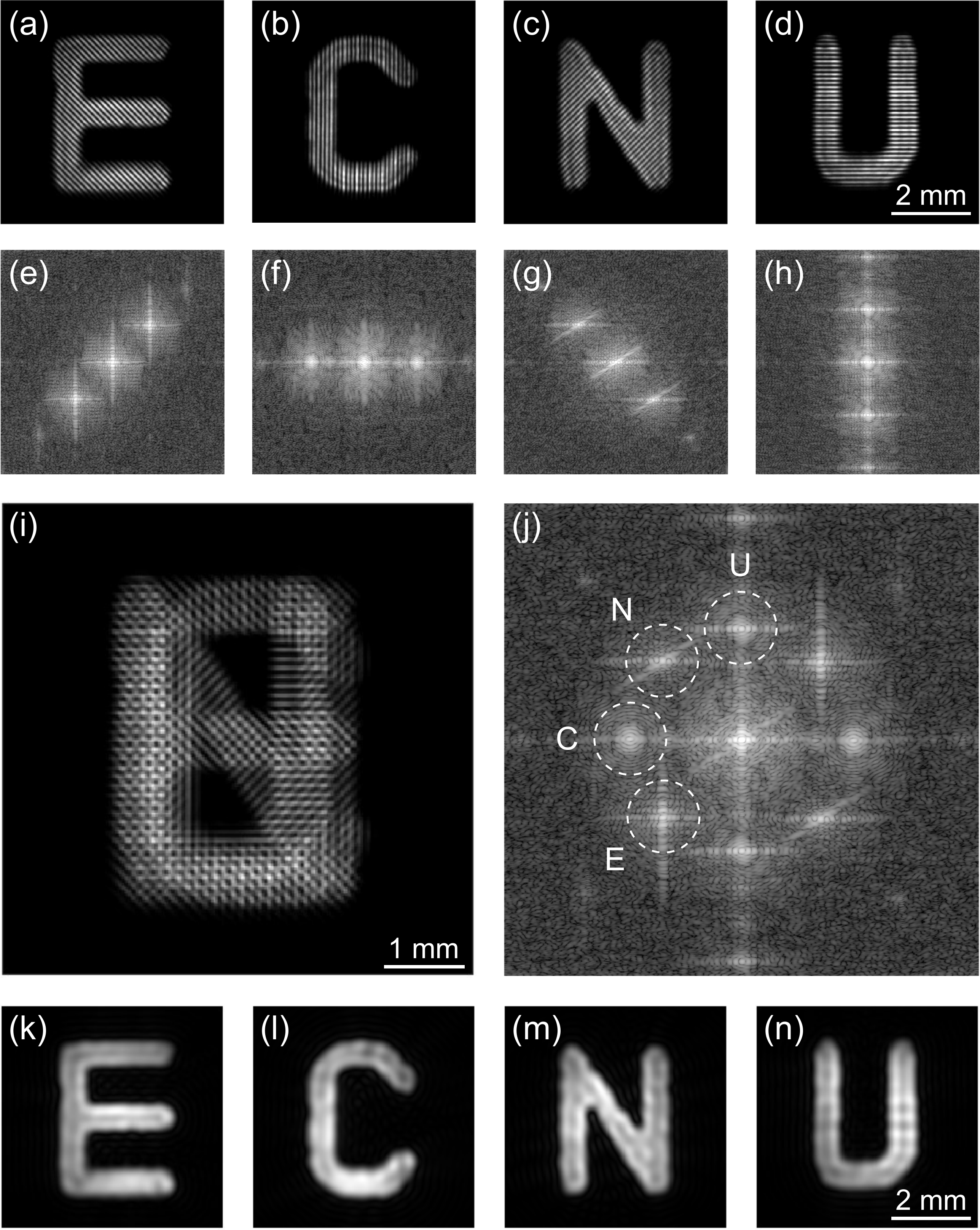}
\caption{Performance characterization of the MIR multiplexed imaging. (a-b) Four object images in the presence of distinct spatial modulation patterns with orientation angles of 135$^\circ$, 90$^\circ$, 45$^\circ$, and 0$^\circ$, respectively. (e-h) Corresponding spatial frequency distributions at the Fourier plane. (i) Integrated image with four modulated frames. (j) Corresponding Fourier frequency distribution. (k-n) Reconstructed images for the four targeted letters.}
\label{fig3}
\end{figure*}

In particular, the pump beam is rapidly modulated by a programmed DMD with a maximum switching rate of 10 kHz. As a result, a sequence of stripe patterns can be prepared with various spatial periods and orientation angles, which allow to upshift the frequency components of time-varying object images into distinct regions at the Fourier domain \cite{Ehn2017LSA, Gragston2018OE}. The time-multiplexed nonlinear structured detection is the key to acquiring both the spatial and temporal information in a single intensity image, in marked contrast to previous upconversion imagers based on a static Gaussian pump \cite{Junaid2019Optica, Huang2022NC, Fang2024NC}. The upconversion beam after the optical modulation passes through a series of spectral filters before being recorded by a silicon camera. The camera permits a frame rate of 1 kHz for a large spatial format with 1024$\times$1024 pixels. The captured image can be demultiplexed to recover the transient frames, which effectively improves the frame rate. More details on the experimental setup and configuration settings are presented in Supplementary Note 1.

\section{Experimental Results}
\subsection{Characterization of imaging performance}
Now we start to characterize the MIR imaging system with a USAF-1951 resolution target. The object image is relayed by two lenses into the camera. The scaling factor from the MIR image to the SFG replica is calculated to be about 3.1 according to the formula $\mathcal{M} =  z_2 /  z_1 \times  z_4 /  z_3$ \cite{Barh2019AOP}, where $z_{1-4}$ are distances between the conjugate planes and the lenses. Note that the scaling factor is independent of the incident wavelength, which contrasts to the spectral radial dispersion in the 4f-based upconversion imaging configuration \cite{Fang2024NC}. The recorded upconversion image is displayed in Fig. \ref{fig2}(a) in the absence of the spatial modulation on the pump, when the DMD is operated at the all-on state. Figure \ref{fig2}(b) presents the cross-sections of the horizontal and vertical lines of the first element in the third group, corresponding to a specified bar width of 62.5 $\mu$m. The uneven heights for the three resolved peaks are ascribed to the inhomogeneous intensity distribution of the Gaussian MIR illumination source. Figure \ref{fig2}(c) presents the SFG image for a pump modulation along 135$^\circ$ within the nonlinear crystal, which shows off-diagonal fringes with a linewidth of 42 $\mu$m. Note that the observed spatial resolution for the structured pattern is higher than that for the line pairs of test target in the object plane. This is due to the fact that the pump pattern is directly mapped onto the scaled-down MIR image in the nonlinear crystal, without suffering from the spatial frequency filtering of the entrance aperture. The modulation period is close to the resolving power of the upconversion imager, which is manifested in the frequency distribution at the Fourier domain as shown in Fig. \ref{fig2}(d). The central spectral components corresponding to the resolvable frequencies are shifted to a series of harmonic regions that are almost connected with each other. As expected, the harmonics decrease in magnitude as the order increases. To this end, the first-order harmonic component is selected via a band-pass filter, and then shifted back to the center of the Fourier domain to remove the spatial modulation. The demodulated image is given in Fig. \ref{fig2}(e), where the spatial resolution is maintained as verified by the cross-sections along the two orthogonal directions in Fig. \ref{fig2}(f).

\begin{figure*}[t!]
	\includegraphics[width=0.69 \textwidth]{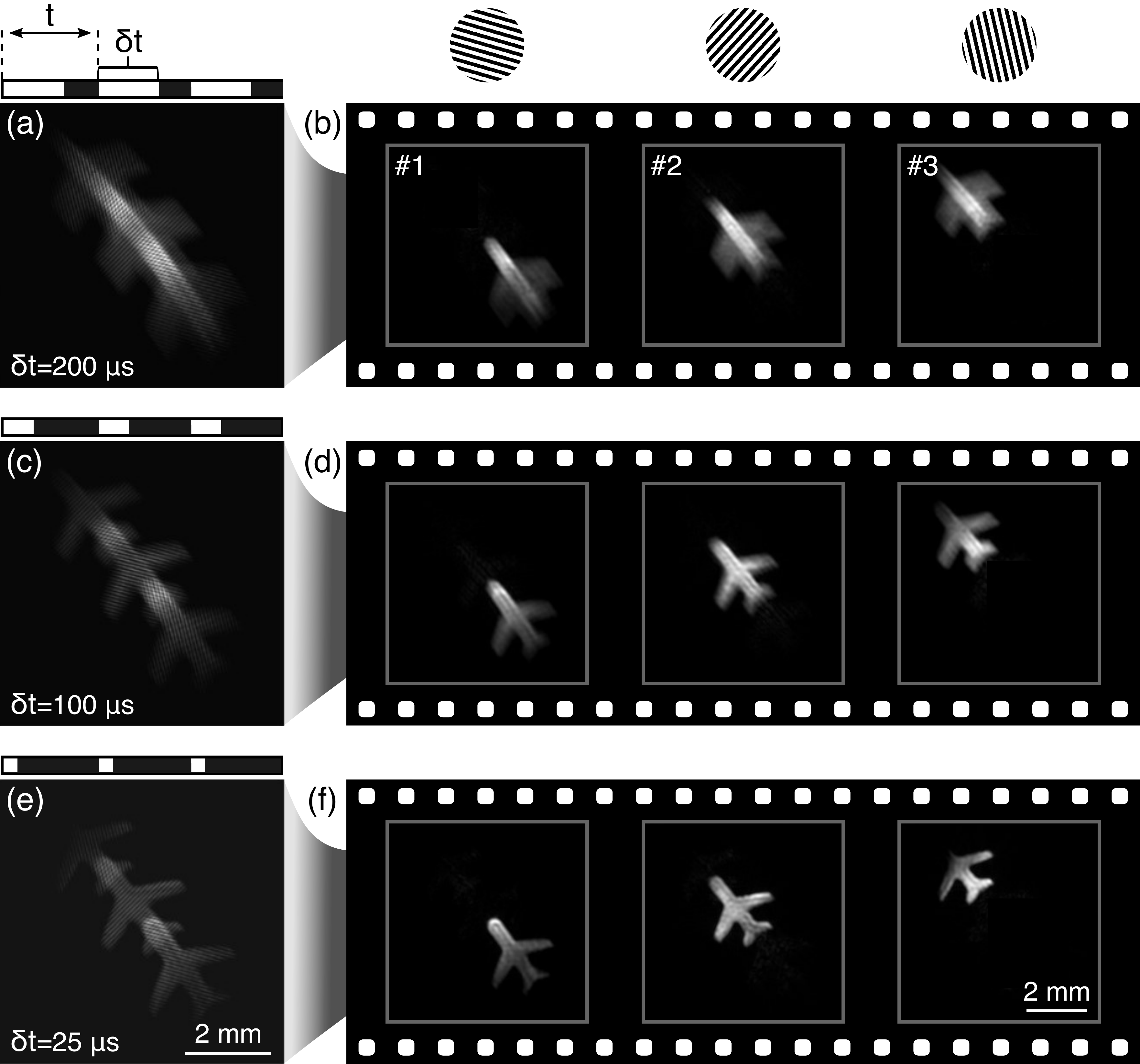}
	\caption{MIR stroboscopic photography based on the nonlinear structured gating. The exposure time for the camera is set to be 1 ms, corresponding to the frame rate of 1 kHz. The exposure window is temporally segmented into three sections with a duration of $t$=1/3 ms. For the image capturing at each section, the pump duration $\delta t$ can be controlled to provide an all-optical shutter. (a, c, e) Recorded upconversion images for $\delta t$ = 200, 100, and 25 $\mu$s, respectively. (b, d, f) Recovered sub-exposure images related to the three modulation patterns. Note that the blurring effect in the demultiplexed images in (b) is due to the relatively long gating time instead of the reconstruction procedure. The line speed of the moving object is about 5.5 $m/s$.}
	\label{fig4}
\end{figure*}

\subsection{Nonlinear multiplexed imaging}
Next, we turn to investigate the performance of the MIR multiplexed imaging. In this case, transmission masks with four letters ``E", ``C", ``N", and``U" are used to prepare the object images, which are modulated respectively by four spatial patterns aligned at 135$^\circ$, 90$^\circ$, 45$^\circ$, and 0$^\circ$, as presented in Figs. \ref{fig3}(a-d). Figures \ref{fig3}(e-h) give the Fourier-transformed distributions, which show the corresponding spatial spectra and shifted harmonics along different directions. The four object images are superposed to form a multiplexed image as depicted in Fig. \ref{fig3}(i). From the corresponding Fourier spectrum shown in Fig. \ref{fig3}(j), we can see that the spectral information for the four letters is distinctly located at a ring of harmonic regions. Each letter is identified by a dashed circle. Figures \ref{fig3}(k-n) present the reconstructed images after the demultiplexing operation, which indicate the competence of the multiplexed imaging. Note that the circle diameter is related to the filtering bandwidth for the bandpass. The filter bandwidth should be as large as possible to retain more spectral components for optimizing the spatial resolution, while avoiding the overlap between adjacent filtered regions. More discussion about the spatial resolution depending on the filter size is given in Supplementary Note 4.

\begin{figure*}[t!]
	\includegraphics[width=0.75 \textwidth]{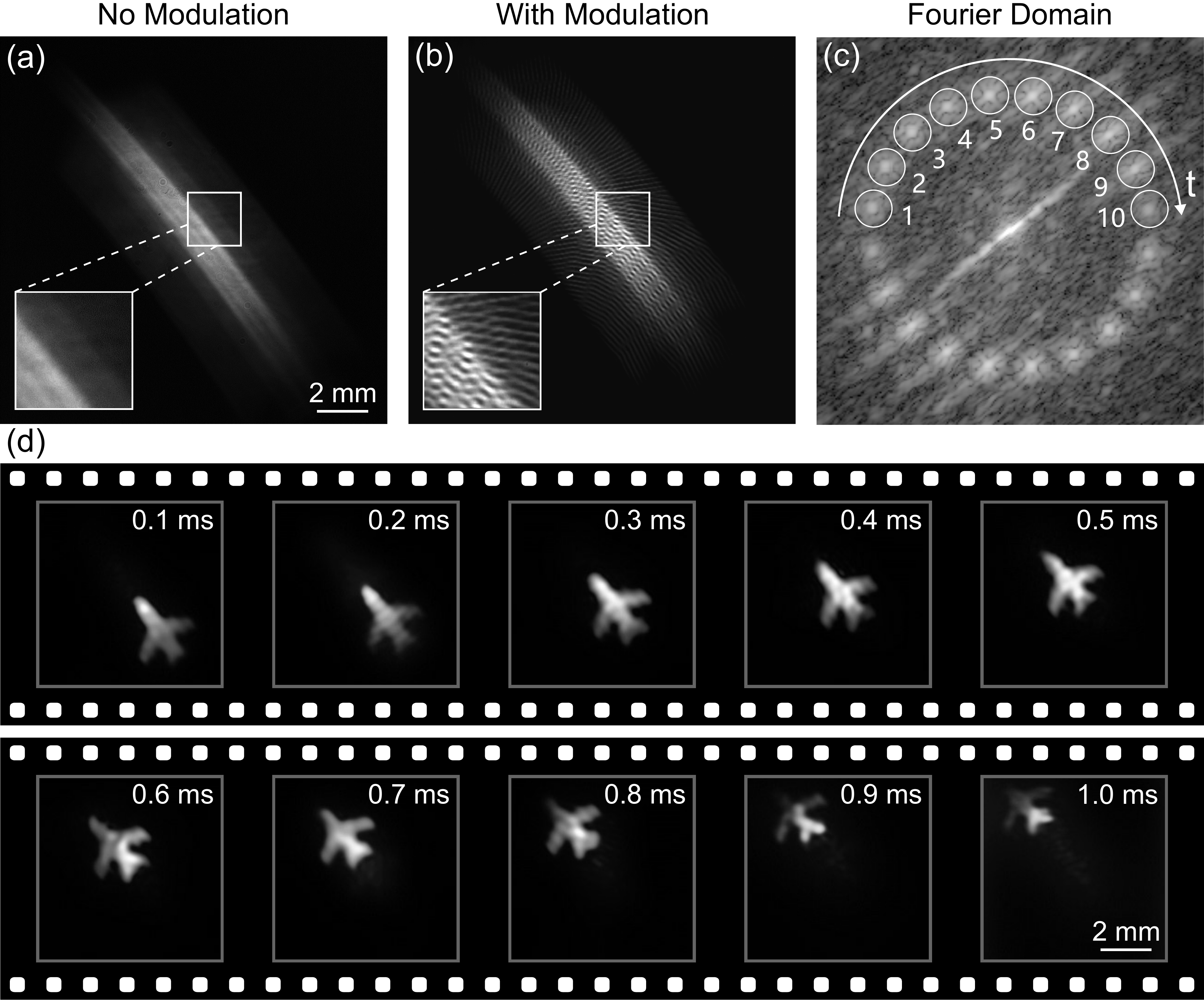}
	\caption{High-speed MIR videography. (a) Captured snapshot in a cumulative time of 1 ms in the absence of the spatial modulation, which shows blurring traces for the moving object. (b) Recorded single-shot image with the dynamic structured modulation, which exhibits multiplexed fringes as shown in the inset. (c) Corresponding spatial frequency distribution at the Fourier plane, which indicates ten image replicas in distinct regions at an upshifted frequency. (d) Reconstructed sequence of the sub-exposure transient images, which illustrate the temporal evolution of the fast moving object. Note that the continuous-stream MIR imaging at a frame rate of 10 kHz is presented in Supplementary Video 1.}
	\label{fig5}
\end{figure*}

\subsection{MIR stroboscopic photography}
In the following, we proceed to implement the MIR stroboscopic photography based on the nonlinear structured detection. Conventionally, the stroboscopic imaging technique uses short and rapid flashes of light at regular intervals to capture fast-moving subjects or events. This allows for effectively ``freezing" the action in a series of overlapping images on a single frame, thus permitting the visualization of motion over time that is otherwise too quick to see with the naked eye. Similar to the illumination scheme based on the burst of pulsed light from the strobe, the detection strategy can equivalently perform the required temporal modulation by a fast switching shutter. However, the switching time of common cameras is typically limited by the speed of the mechanical or electronic shutter. In our experiment, the nonlinear optical modulation offers an alternation to perform the high-speed repetitive gating operation within a single exposure window. Here, the object target is a rapidly rotating wheel engraved with an airplane pattern at a speed of 1380 revolutions per minute (rpm), corresponding to a line velocity about 5.5 m/s. The DMD operates at a modulation frequency of 3 kHz, while the illumination time $\delta t$ for each pattern can be set much shorter than the pattern switching period $t$. The exposure time for the camera is set to be 1 ms, which can accommodate three gating patterns with different spatial fringes. Figure \ref{fig4}(a) shows the captured multiplexed image at $\delta t$ = 200 $\mu$s, which exhibits a clear blurring effect due to the object displacement at each optical gating time. Such a blurring is also manifested in the three demultiplexed frames as given in Fig. \ref{fig4}(b). As expected, the blurring effect can be mitigated by using a shorter gating time, as verified by the observations in Figs. \ref{fig4}(c) and (d). Particularly, the MIR stroboscopic photography is unambiguously demonstrated at $\delta t$ = 25 $\mu$s as illustrated in Fig. \ref{fig4}(e). In contrast to traditional stroboscope, our approach not only allows one to capture a stack of motionless frames at different moments, but also offers the possibility to separate these spatially overlapped images. Figure \ref{fig4}(f) presents the three recovered transient frames to reveal the time-varying events. We note that the ultrafast gating time beyond $\mu$s can be accessed with faster pattern generation \cite{Kilcullen2022NC} or path-division configuration \cite{Ehn2017LSA}, which would offer an enabler in observing highly dynamic infrared scenes.

\subsection{High-speed MIR videography}
Finally, we demonstrate the high-speed MIR videography via the Fourier-domain multiplexing. To this end, the DMD is running at the maximum modulation frequency of 10 kHz. The camera works at a frame rate of 1 kHz with an acquisition time of 1 ms, which allows for integrating ten sub-exposure images under a sequence of predefined optical pumping patterns. The rotation speed of the target is the same as the aforementioned setting, which has been optimized to ensure that the object traverses the entire field of view. As a comparison, Fig. \ref{fig5}(a) presents the recorded single-shot image without the spatial modulation. Due to the limited frame rate of the camera, contour information of the object can barely be discerned from the image. Figure \ref{fig5}(b) shows the multiplexed image with the integration of ten modulation fringes, which exhibits a complex intensity pattern for maintaining spatiotemporal information of the dynamic target. The corresponding frequency distribution at the Fourier domain is presented in Fig. \ref{fig5}(c), which clearly reveals the ten pairs of first-order harmonic components along the directions from 0$^\circ$ to 162$^\circ$ with an interval of 18$^\circ$. Figure \ref{fig5}(d) shows the reconstructed sub-exposure frames in a chronological order during the 1-ms integration period, as indicated by the demultiplexed moments in the upper-right corner of each frame. Thanks to the high-speed nonlinear structured multiplexing, the effective frame rate of the camera is substantially enhanced to 10 kHz by an order of magnitude, hence enabling precise visualization of the object's motion that is otherwise obscured by the blurring effect. Moreover, the snapshot videography favors the continuous-stream capturing of multiplexed images, which allows one to record the real-time video of the transient scene at a high frame rate and a long acquisition sequence as given in Supplementary Video 1.

We note that the frame rate can be further boosted by approaching the maximum multiplexing condition, where maximum number of frames with optimized spatial resolutions are obtained by fully occupying the frequency domain up to the diffraction limit circle of the optical imaging system \cite{Ehn2017LSA}. In this case, the structured pump patterns are designated with varying angles and cycle periods to shift the sequence of exposures to various spatial locations at the Fourier plane \cite{Gragston2018OE, Gragston2018AO}. There is a theoretical limit for the multiplexing single exposure, which is determined by the operation bandwidth of the frequency filter and numerical aperture of the imaging system (see Supplementary Note 5). Indeed, the spatial resolution of time-multiplexing images is reduced and controlled by the band-pass filter, \textit{i.e.}, the size of the circle in Fig. \ref{fig5}(c), which is inherently determined by the uncertainty principle of Fourier transformation \cite{Khan2018SPIC}. In practice, it is essential to design proper multiplexing patterns to balance the need between the desired spatial resolution and the frame-rate scaling factor. In many scenarios of high-speed image analyses, the information is mostly concentrated at the central Fourier region with low spatial frequencies. Hence, the band-pass filter only slightly reduces the spatial resolution, which can still satisfy the requirement for the model development and validations in the related fields pertinent to investigating turbulence, combustion, or aerodynamics \cite{Gragston2018AO, Cai2021RSI}.

\section{Discussions and conclusion}
The presented architecture is inspired by the computational imaging modality termed as frequency recognition algorithm for multiple exposures (FRAME) \cite{Ehn2017LSA, Gragston2018OE, Gragston2018AO, Cai2021RSI, Khan2018SPIC}. However, previous demonstrations are restricted in the visible or NIR regions due to the accessibility of sensitive and fast imagers. Recently, the FRAME technique has been extended to demonstrate a single-shot ultrafast terahertz (THz) photography, albeit with a low sequence depth due to the free-space multiplexing configuration and a limited detection sensitivity due to the inefficient third-order nonlinear process \cite{Dong2023NC}. In these reported schemes, the illumination beam is only subject to the spatial modulation without the wavelength conversion. In contrast, the nonlinear structured imaging scheme presented here provides a conceptually novel approach to simultaneously realize the spatial modulation and spectral conversion based on the second-order SFG, which allows high-fidelity masking and high detection sensitivity for the MIR radiation. Moreover, the interleaving operation with demultiplexed frames facilitates a continuous-stream MIR imaging at high speed and high definition, which outperforms current state-of-the-art MIR cameras.

As summarized in Supplementary Note 6, we have presented a thorough comparison among various representative schemes for wide-field MIR imaging systems. Historically, the performances of the direct MIR imagers are hindered by the slow refreshing rate and limited detection sensitivity, especially for large spatial arrays with high pixel counts \cite{Razeghi2014RPP}. Indeed, fast MIR cameras based on microbolometers or narrow-bandgap semiconductors are typically restricted to small focal plane arrays. For instance, the frame rate of state-of-the-art InSb-based imagers reaches to 1.5 kHz for 640$\times$512 pixels, which dramatically decreases to 180 Hz for 1280$\times$1024 pixels \cite{Telops}. In the context, the upconversion strategy has been adopted to leverage the high-performance cameras in the visible or NIR regions. Specifically, high-speed MIR imaging based on the non-degenerate TPA effect has been demonstrated with a frame rate of 100 Hz for a spatial resolution of 1024$\times$1024 pixels based on the InGaAs camera \cite{Potma2021APLP}. Moreover, the use of more efficient SFG process favors the substantial reduction of exposure time for obtaining high-contrast images, which allows a frame rate up to the kHz-level with a Si-based camera \cite{Huang2022NC, Fang2024NC, Junaid2019Optica, Zhao2023NC}. However, the ultimate speed for the image acquisition with a high-definition pixel format is restricted by the intrinsic frame rate of the involved camera. In our work, the proposed multiplexing imaging strategy makes it possible to overcome the physical limit, which allows us to achieve a record-high frame rate up to 10 kHz for a megapixel spatial format.

In contrast to previous approaches of the coded optical imaging \cite{Yao2024LPR, Liang2024Book}, the implemented scheme offers several distinct features. First, the coding operation is applied on the pump beam, instead of on the MIR signal beam. Such an indirect optical modulation can mitigate the bottleneck in high-fidelity spatial modulation at longer wavelengths \cite{Edgar2019NP}. Second, the spatially encoded MIR beam is spectrally converted into the replica at the NIR wavelength, which permits the use of fast and sensitive silicon-based cameras \cite{Wang2023NC} to access superior imaging performances beyond the current MIR sensors \cite{Razeghi2014RPP}. Third, the coding patterns are based on periodical stripes at different orientations, which favors fast and stable image reconstruction through simple operations based on frequency filtering and Fourier transformation \cite{Ehn2017LSA, Gragston2018OE, Gragston2018AO}. Fourth, the presented nonlinear structured detection approach offers the ability to apply the modulation on the imaging side, which is
beneficial to observe natural luminous objects via the passive thermal imaging \cite{Dam2012NP}.

To go beyond the achieved performance, the compression ratio for the imaging multiplexing can be further improved by enlarging the numerical aperture of the optical system, performing compressive-sensing measurements, and adopting more advanced deep-learning reconstruction algorithms \cite{Wang2024LAM}. Accordingly, a faster spatial modulation should be adapted, for instance resorting to the approaches based on swept aggregate patterns \cite{Kilcullen2022NC}, which could permit a high-speed imaging at the megahertz level. Furthermore, time-varying pump patterns with picosecond intervals can be realized by resorting to the path-division configuration, which open up possibilities for single-shot MIR ultrafast photography \cite{Ehn2017LSA} and snapshot MIR depth imaging \cite{Dong2023NC}. Notably, it is feasible to extend the presented approach into longer infrared or terahertz regions \cite{Rodrigo2021LPR, Fandio2024OL}, where high-speed and high-resolution videography is highly demanded.

In conclusion, we have implemented a high-speed and high-definition MIR imaging system, which combines the merits from frequency upconversion detection and coded computational imaging. The presented MIR imager overcomes the intrinsic predicament in simultaneous realization of massive pixels and fast speed, and allows high-throughput continuous image acquisition with a frame rate of 10 kHz and a megapixel spatial format. Such a versatile imaging modality would provide a practical alternative or enabling technique for various applications that require wide-field observation of the involved transient phenomena in a high temporal precision.

\section*{Acknowledgements}
This work was supported by Shanghai Pilot Program for Basic Research (TQ20220104); National Natural Science Foundation of China (62175064, 62235019, 62035005); Innovation Program for Quantum Science and Technology (2023ZD0301000); Natural Science Foundation of Chongqing (CSTB2023NSCQ-JQX0011, CSTB2022TIAD-DEX0036); Shanghai Municipal Science and Technology Major Project (2019SHZDZX01); Fundamental Research Funds for the Central Universities.

\section*{Conflict of Interest}
The authors declare no conflict of interests.

\section*{Supporting Information}
Supporting Information is accompanied to present more details on the data acquisition and processing. 
%available from the Wiley Online Library or from the author.

\section*{Data Availability Statement}
The data that support the findings of this study are available from the corresponding author upon reasonable request.

\section*{Keywords}
mid-infrared imaging, high-speed imaging, computational imaging, frequency upconversion detection

\end{document}